# STYLIZED STATISTICAL FACTS OF INDONESIAN FINANCIAL DATA

Empirical Study of Several Stock Indexes in Indonesia

**Hokky Situngkir**[1]
(hokky@elka.ee.itb.ac.id)
Dept. Computational Sociology
Bandung Fe Institute

**Yohanes Surya**[2]
(yohaness@centrin.net.id)
Dept. Physics
Universitas Pelita Harapan

[1] *Board of Science* Bandung Fe Institute, http://www.geocities.com/quicchote
[2] *Board of Advisory* Bandung Fe Institute

## Abstract

This paper is trying to unveil general statistical characteristic of financial; time series data that is subjected to several financial time series data present in Indonesia, e.g. individual index such as stock price of PT. TELKOM, stock price of PT HM SAMPOERNA, and compiled stock price index (*Jakarta Stock Exchange Index*). Characteristics that we try to analyze are volatility clustering, *truncated Levy distribution*, and multifractality feature. This analysis is directed for further works of research in making Indonesian artificial stock exchange that gives representation of micro structure of stock exchange in Indonesia. This paper is a resume of statistic behavior analyzed in *top-down* to become the ground in starting a more *bottom-up* analysis.

**Keywords:** Indonesia stock exchange, Telkom, HM Sampoerna, stock price index, stock exchange index, volatility clustering, truncated Levy distribution, multifractality.

## 1. Background

There are only few literatures discussing on the statistical characteristics from numbers of financial time series data that we have in Jakarta Stock Exchange (JSX). Furthermore, today analysis of financial economics in Indonesia tends to oversimplify quantitative problems of financial economy using qualitative analysis which in return seems to be frequently very speculative. However, financial data in Indonesia is very specific or unique; henceforth we need to build analysis without ignoring contemporary financial economy analysis that is developing in international-class scientific stages: econophysics with an "instrument" called statistical mechanics.

In advanced of financial economy data analysis, there are several things that in earlier time were developed to approach physics system that is now has largely changed. To day we understand that the characteristic of financial economy data is certainly will be beneficial in micro-simulation approach of financial economy process where the output frequently confusing analysts[1].

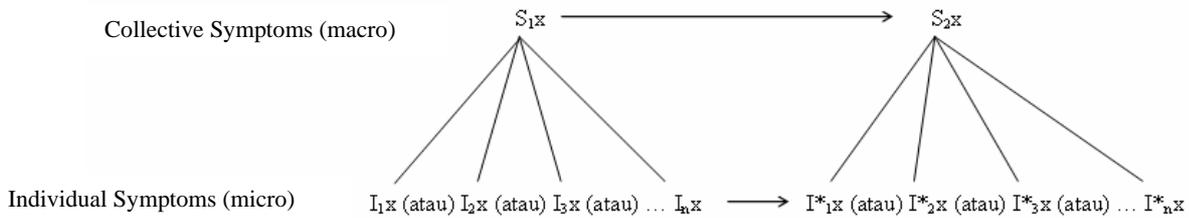

Collective Symptoms (macro)

Individual Symptoms (micro)

**Figure 1**
Financial time series data as a quantitative approach of collective symptom emerge from various individual symptoms causally

As already discussed in Situngkir (2003), social system can be differed into several description levels as objects we will observe. On macro level we can fairly say that, for instance, one collective behavior $S_1x$ causing sociological behavior $S_2x$, with x is a form of community or society that is localized so it ease our analysis. Either $S_1x$ and $S_2x$ should be reminded to occur by "random" interaction of individuals constitute it (let say $I_1x$, $I_2x$, …$I_nx$). The result of collective behavior we analyze using aggregation technique-that in this kind of financial economy system named as price fluctuation, foreign currency, trading volume, etc.

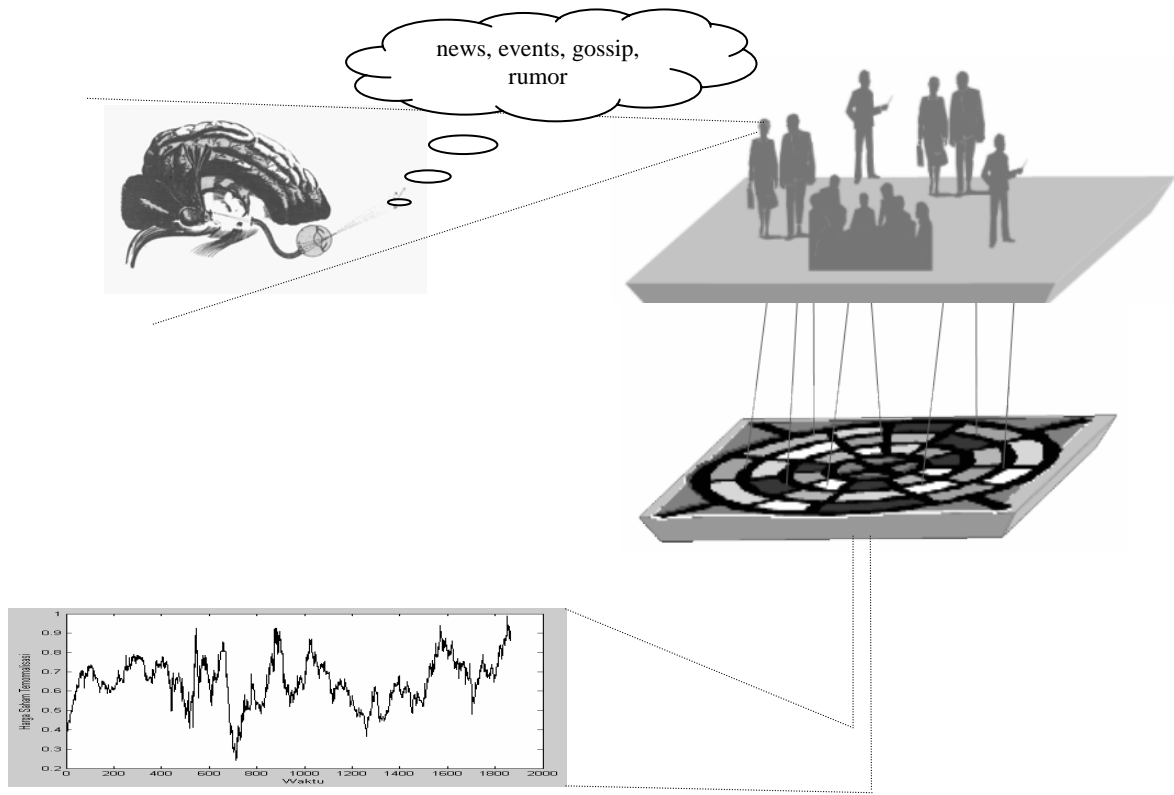

**Figure 2**
The changes of price fluctuation by event, issue, gossip

Statistical analysis commonly view these data as objects of observation without concerning micro impact that they cause – the result thus very often fluctuating changes occur in financial time series data directly related with issue, rumor, gossip, or occurring events. In fact as it is showed in Figure 2, between events and financial time series data, lies long chain which make effort of connecting an event to fluctuation occurs become naïve and false.

---

[1] Further explanation of conventional endeavor that correlates aggregated data with factor that suppose to be micro-analysis can be reviewed in Keen (2002).

This paper offers pictures of general characteristics from statistical financial economy data in Indonesia that we take from Jakarta Stock Exchange. Here is discussed few general characteristics of financial data regards to some statistical mechanics earlier investigation and we will apply them in Indonesia along with meaning of several stated statistical characteristics qualitatively. It is hoped that the paper able to give brief description of statistical behavior of general financial data in Indonesia (mainly in several individual indexes marketed in Jakarta Stock Exchange). In its analysis, this statistical behavior will be subjected to further work i.e. financial micro-simulation and endeavor of forming artificial Jakarta Artificial Stock Exchange[2]. By comprehending general characteristics of aggregative financial economy data, we will possibly work in micro level description and the financial economy data itself. In other words, in micro (agent-based) analysis we will perform, general statistical financial economy data will become our border of how far we can trust that micro-analysis result.

## 2. General Characters of Financial Time series Data

Basically we can see basic characters of financial statistical time series data in Indonesia that is basically have been analyzed by numbers of financial physics references for many fluctuations of financial economy system in United States and Europe (Cont, 1999, Canessa, 2003, Gabaix, et.al., 2003, Farmer, et.al., 2003, and Bouchaud, 1999). This paper is discussing three characters of financial time series data, they are:
- ✓ Characters of Volatility Clustering
- ✓ Characters of Excess Kurtosis and fat tail distribution
- ✓ Multifractality

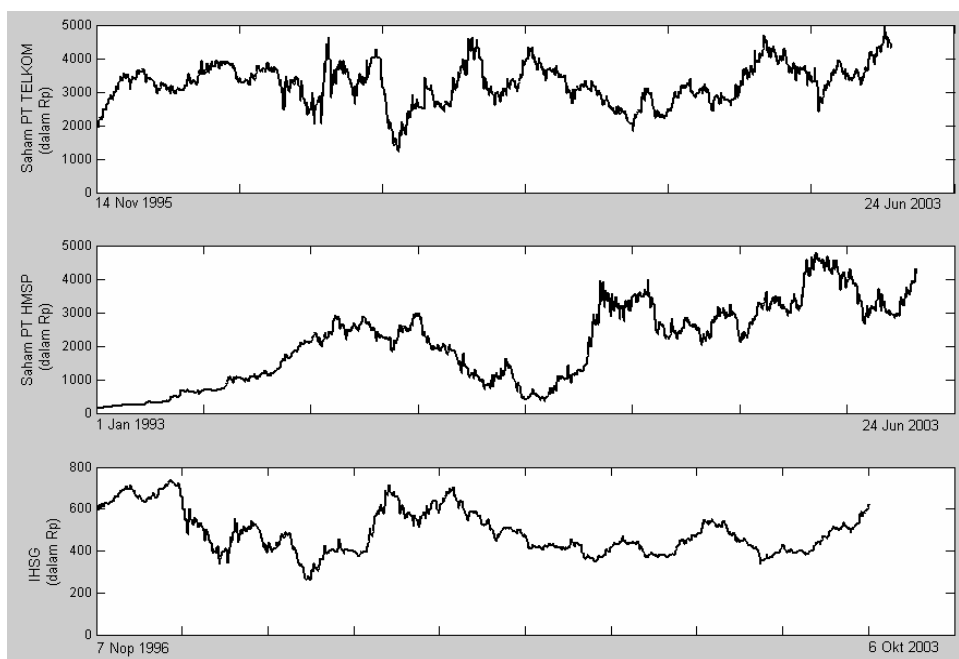

**Figure 3**
Stock price of PT TELKOM, PT HMSP, and IHSG

The above characters have been known as basic characters that are generally prevail in stylized statistical facts of financial time series data.

The chosen financial time series data to be analyzed are individual index data of PT TELKOM, PT HM Sampoerna, and compiled indexes of Jakarta Stock Exchange Index (IHSG). Figure 3 shows fluctuations of each index.

---

[2] This has been designed by Situngkir & Surya (2003b).

In analyzing financial economy data, the center point of our interest is the fluctuation of price that currently occurs. Basically price fluctuation is a variable expressing the up-and-down of price as a causal form of market mechanism happens. Fluctuation has been tempting analysts' attention up to now since there are so many definitions given in representing price fluctuation.

Commonly, the direct and traditional definition of price fluctuation is simply the changes of price:

$$\Delta p_t = h_t - h_{t-1} \qquad \ldots(1)$$

or technically the change of price at time **t** noted by $\Delta p_t$, is the difference of current price ($h_t$) with the previous price ($h_{t-1}$); where **t** is series of time which can have unit of second, day, month, to years. Furthermore, the approach for price fluctuation is the relative change or *return* that commonly defined as *continuously compounded return* or *log-return*, that is:

$$z_t = \ln p_t - \ln p_{t-1} \qquad \ldots(2)$$

This is the dimension that nowadays frequently used in numbers of advanced analysis of financial economy data that in practice is really large numbers. Through the use of this parameter, it certainly will be easier to analyze them as a form of analysis and processing of a ultra-high frequency signals. Figure 5 shows absolute value of the volatility of the three financial time series data that we will discuss further.

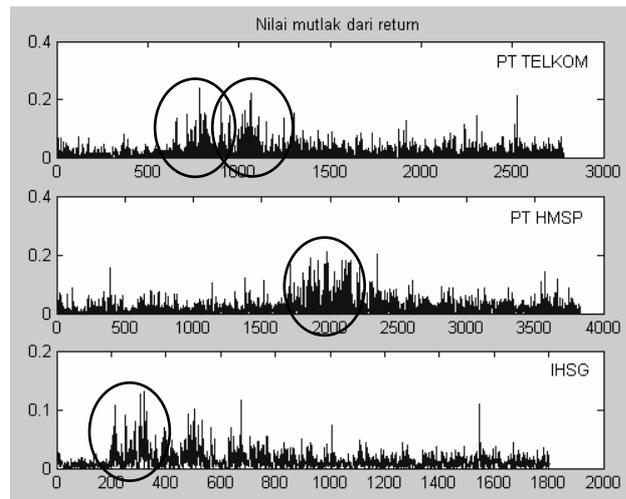

**Figure 4**
Absolute value of each index *return*. Notice the volatility clustering occurs: the change of price value tends to cluster: high change with large change and the low change with small change.

## 3. Volatility Clustering

Volatility is the terminology of sensitivity of a financial time series data. Commonly this dimension is expressed as deviation standard from the speed of changing compounds of financial time series data in ARCH (*Autoregressive Conditional Heteroskedascity*) analysis or the common form (*Generalized* ARCH) in certain variation. In short, volatility is a measure of uncertainty of the financial time series data or possible risk that is commonly faced by investors in stock trading.

Plainly, several references (Castiglione, 2001:74) describe volatility as amplitude of increment of time series data that expressing volatility as absolute price from the value of *return* ($v_t = \sqrt{r_t^2}$). In accordance with fluctuation movement of price as Brownian motion, in further volatility can be regard as process coefficient of stochastic Brownian motion (Baxter & Rennie, 1999:55).

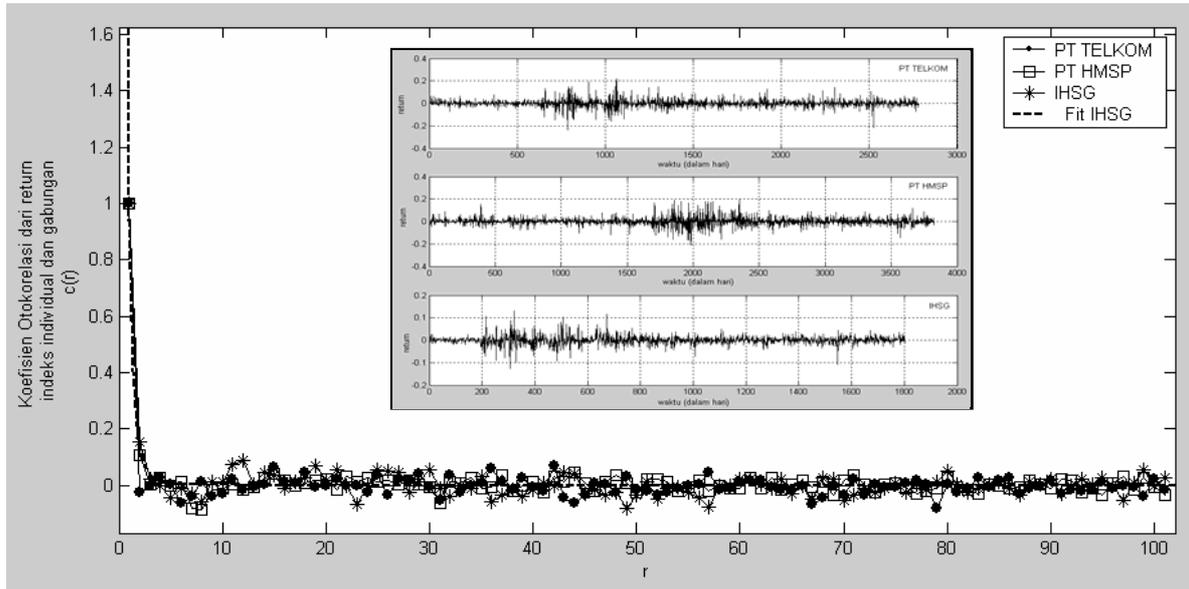

**Figure 5**
Autocorrelation of *return* of each price fluctuation index that decreasing. It is shown that *sampling* distribution index of IHSG that is fitted exhibiting the declining of *power-law* with order coefficient of about 2,911. Inset shows the *return* plot of each that we try to find its autocorrelation coefficient.

The symptom of volatility clustering is seen as positive autocorrelation function and declining so it is reaching zero. For autocorrelation function of several stock data in Indonesia, we can see in (Situngkir & Surya, 2003b), which is stated by Nist-Sematech (2003) that if the data shown in time series of $y_i$ with $i = 1, 2, 3, ...$, thus the autocorrelation coefficient can be written as:

$$r_k = \frac{\sum_{i=1}^{n-k}(y_{i+k} - \bar{y}_{i+k})(y_i - \bar{y}_k)}{\sqrt{\sum_{i=1}^{n}(y_i - \bar{y}_i)^2 \times \sum_{i=1}^{n}(y_{i+k} - \bar{y}_{i+k})^2}} \quad ...(3)$$

where $r_k$ is the autocorrelation of $y_i$ and $y_{i+k}$. Autocorrelation of several samples of data forming distribution of around $k$ is commonly called *sampling* distribution autocorrelation.

From Figure 5, we can find out that autocorrelation function of the three indexes of stock value analyzed have a similarity – where in *sampling* distribution autocorrelation from compiled index IHSG can be approximated as *power-law* distribution with order coefficient of 2,911. This means that distribution index (either individual or grouped) of financial economy data in Indonesia also follows character of the universal from the financial time series data. In term of this financial time series behavior, we certainly will see in microstructure view of it as is explained in (Gaunersdorfer & Hommes, 2000) for several financial economy time series data of S&P500.

## 4. Levy Distribution & Scaling Behavior

Another very unique character in financial time series data is the distribution of financial data. The character is following non-Gaussian distribution. Mantegna & Stanley (2002:63-75) showed how behavior of distribution of financial economy time series data generally has below characters:
- ✓ Nearly symmetric
- ✓ Strongly Leptokurtic
- ✓ Assigned with profile of non-Gaussian distribution for small change of index.

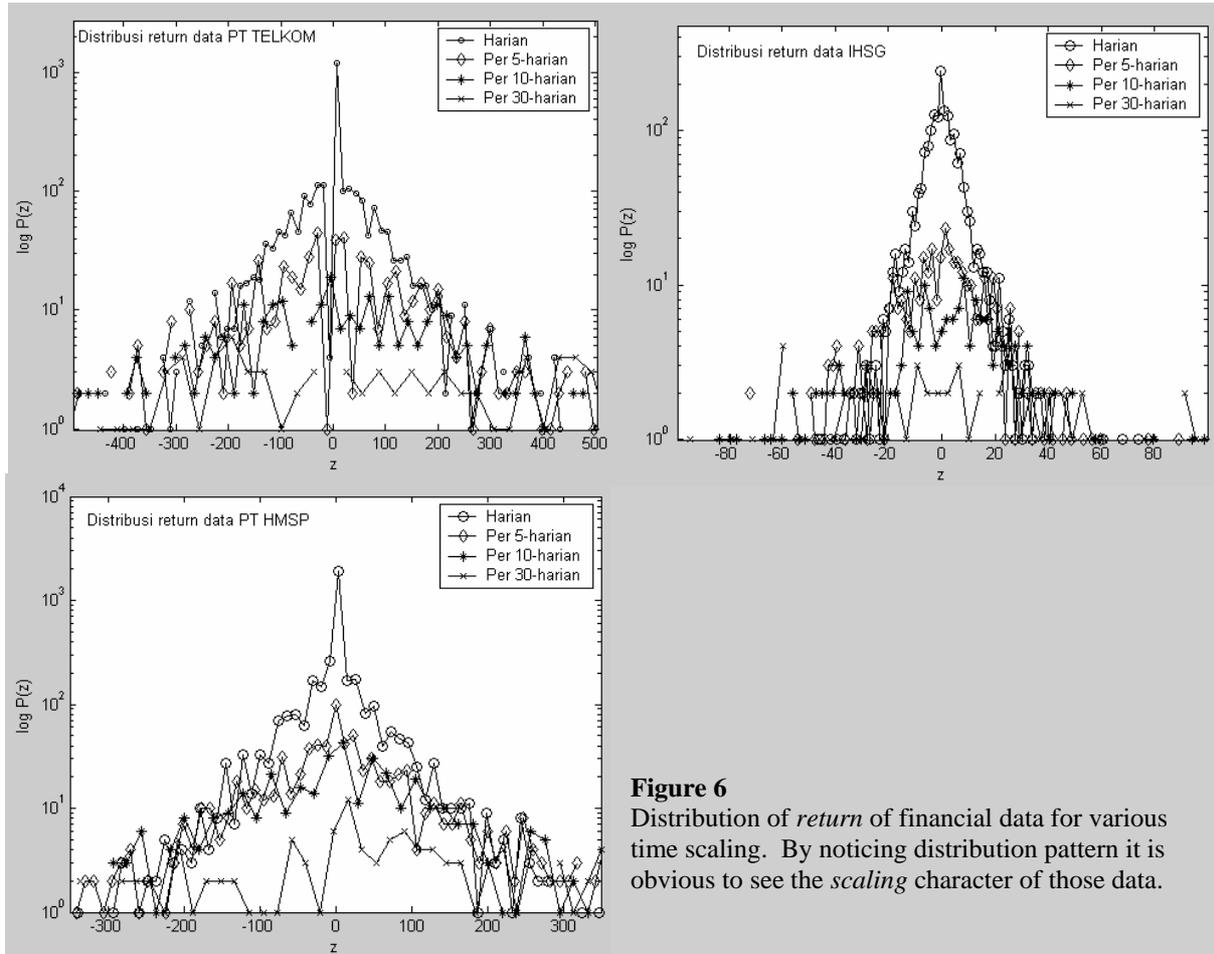

**Figure 6**
Distribution of *return* of financial data for various time scaling. By noticing distribution pattern it is obvious to see the *scaling* character of those data.

The three characters above is viewed obviously in the description of Figure 6. Theoretical exploration about this can be view in Situngkir & Surya (2003a). Several distribution characters from the data we have processed here described by table 1 where the value of several dimension of things used as parameters of a data distribution characters, i.e. standard deviation, *skewness*, and kurtosis. *Skewness* can be understood as asymmetrical measure of the data around sample average. If *skewness* is negative, thus the data tend to the left and vice versa if it is positive it tends to the right. *Skewness* value from the truly symmetric distribution (e.g. normal distribution) is zero. *Skewness* in this paper is defined as:

$$y = \frac{\mu_3}{\sigma^3} \qquad \ldots(4)$$

with $\mu_3 = \langle X - \sigma \rangle^3$ which is the half third moment and $\sigma$ as deviation standard[3]. In the other side, kurtosis is a measure of data tendency that lies beyond distribution. Kurtosis of the normal distribution is 3, means that if kurtosis larger than 3 the samples of data tend to lie beyond normal distribution while if kurtosis smaller than 3, they tend to be 'inside' the normal distribution. Kurtosis in this paper defined as:

$$k = \frac{\mu_4}{\sigma^4} \qquad \ldots(5)$$

with $\mu_4 = \langle X - \sigma \rangle^4$ is the forth half moment. Data distribution of financial time series data, as described in Figure 6 showing density of probability mass in the "tail" and "head" distribution that lie beyond the cope of normal distribution. This is what we called with the expression of 'leptokurtosis' or named *fat tails* from distribution of *return* value of financial time series data. Leptokurtic distribution assigned with tight maximum value yet it is large in value, and in return the distribution tail is fatter than Gaussian distribution tail.

We can obviously see in table 6 that each coefficient value stated above expressing the distributional tendency of financial economy data in Indonesia. We will discuss about this more detail at the end of the paper.

**Table 1**
**Resume of statistical distribution of the *return* of the data used in the experiment reported here**

|  | PT TELKOM | PT HMSP | IHSG |
|---|---|---|---|
| **Number of Data (daily)** | 2779 | 3826 | 1802 |
| **Deviation standard** | 0.0292 | 0.0282 | 0.0191 |
| **Kurtosis** | 13.1154 | 13.5568 | 9.9499 |
| ***Skewness*** | 0.1439 | 0.3977 | 0.1458 |
| **Parameter $\alpha$** | 0.4510 ± 0.0226 | 0.683 ± 0.0342 | 1.479 ± 0.074 |

Basically from Figure 6 we can see carefully that the return values of our financial data behave non-Gaussian character for various time-scaling. This character is called by 'scaling' behavior. This is interesting since the exhibited of the behavior of Levy distribution in our financial economy data as *power-law* distribution (Mandelbrot, 1963). Through noticing Lévy *power-law* distribution model in invariant scale (Castiglione, 2001:71-73) we can shortly defined:

$$P(z) \sim |z|^{-(1+\alpha)} \qquad \ldots(6)$$

as we will get $\alpha$ for each index as attached in table 1.

More specifically, Mantegna & Stanley (2002) and Paul & Bashnagel (1999:122-125) exhibiting that the scaling behavior eventually falls down by the rise of interval value used – where time series eventually return to the Gaussian distribution regime. Unfortunately, this cannot be performed in this paper for the lack of the data; that means the stable Levy distribution with clear character of scaling with its infinite variance, basically can not generally describe the financial data distribution. As it has been described in earlier works (Situngkir & Surya, 2003a), financial distribution will always have finite variance and the distribution that able to describe it is *truncated Levy distribution*. The cross-over from Levy truncated distribution to Gaussian distribution is named by the fall of scale or the disappearance of scaling character from time series data distribution.

## 5. Multifractality of stock Indexes in Indonesia

The last character we are going to analyze from our financial economy data that we chose is the multifractality in the available time series data. This character is basically the

---
[3] Symbol $\langle \ \rangle$ representing average value.

advanced form or the consequence of the previous character, i.e. '*scaling*' behavior. In Simple way, we can understand scaling behavior as the result of the transform from *power-law* distribution to financial time series data distribution. *Scaling* character make a condition where the *return* function from price fluctuation be made as:

$$z_\tau(t) = \ln\left(\frac{p(t+\tau)}{p(t)}\right) \qquad \ldots(7)$$

where $\tau$ as certain time scale, may be re-scale for the left factor of $\eta(\tau)$ that occupies:

$$P(z_\tau) = \frac{1}{\eta(\tau)} \Phi\left(\frac{z_\tau}{\eta(\tau)}\right) \qquad \ldots(8)$$

where

$\Phi(x)$ is *scaling* function which is independent over time, self-similarity function of $\eta(\tau) = \tau^H$ with $H = 0.5$ for Gaussian process and $H = \frac{1}{\alpha}$ for Levy process (Iori, 2000). As we have explained previously, in accordance with Center Limit Theorem, the bigger the value $\tau$ we use, thus the scaling character is gradually falling and data distribution is taken over by the Gaussian regime (Mantegna & Stanley, 2002:64-65).

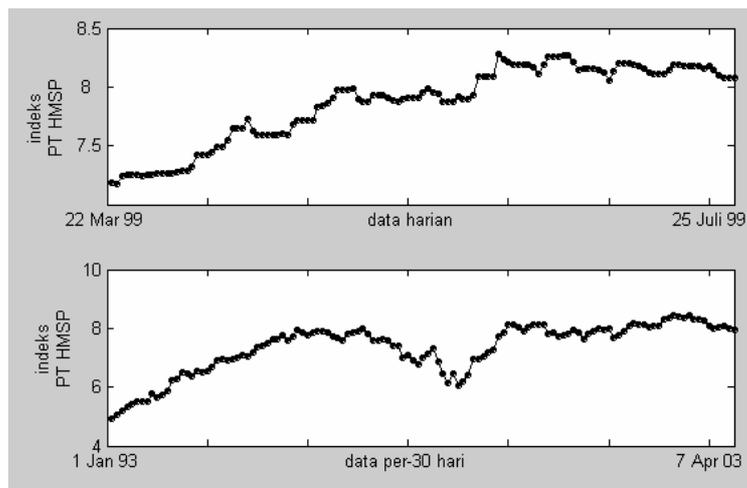

**Figure 7**
Multifractality character of individual index of PT HMSP

In the previous work (Hariadi & Surya, 2003) it has already been explained that according to computation using Hurst exponent, Indonesia financial economy system has multifractality character. Stock price index of PT HMSP is computed to have fractal dimension of 1.7627 while PT TELKOM has fractal dimension of 1.7147. It is an interesting thing to be developed in further correlation of multifractality of stock index in Indonesia. Figure 7 shows multifractality character emerge from stock index of PT HMSP where daily data of March 22nd, 1999 up to July 25th, 1999 has similarity with price index per 30-days for the date of January 1st, 1993 to April 7th, 2003. Similarity we found having a trend character for each increment of index price fluctuation, this is due to the minimum data that we can processed in this discourse of multifractality.

The same is true shown by Figure 8, where certain stock index data per 10-days has similarity with particular part of those daily data of PT TELKOM.

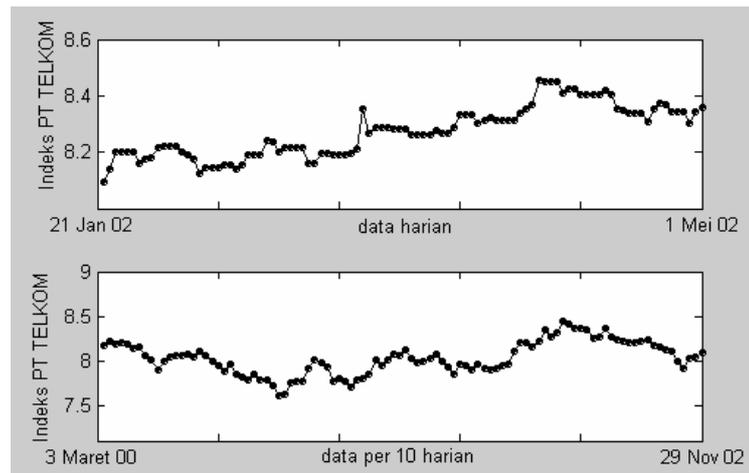

**Figure 8**
Multifractality character in individual index of PT TELKOM

## 6. Few Discussions & Conclusion

The weakness presented in the previous sections is the poor of data thus we do not have ability to proceed giving answers to the following questions:

- ✓ How far Levy distribution regime can stand and replaced by Gaussian regime by the rising of time increment ($\tau$)? Or when is the exact time the falling of scaling behavior in the return distribution of stock indexes in Indonesia?
- ✓ In what location multifractality that lies in index price fluctuation (either individual or collective) in Indonesia?

Both questions above certainly will give new horizon of viewpoints of market price fluctuation we commonly meet in stock exchange in Indonesia. Yet, beyond the weakness we may find in our analysis that financial economy time series in Indonesia much or less is following the stylized behavior of financial economy data that mostly referred by several world-wide analysts.

It is an interesting discussion we have found on the autocorrelation function of individual indexes of (PT HMSP) and (PT TELKOM) and compiled indexes IHSG. In Figure 5 we have seen how autocorrelation function of the three indexes is really close each other in term of fitting plot of autocorrelation function of IHSG (which following *power-law* distribution sample) and plain that they are able to picture the other second individual stock autocorrelation function – this sufficiently explain the influence of the both individual indexes over IHSG as a whole, as two of many large company indexes in stock exchange.

Further works about the exploration of statistical distribution character of present time series data will be very dependent on data supply we can analyze. Yet, in frame of starting analysis work of market micro-structure, what has been described in this paper surely has given a picture of the three financial time series data characters in Indonesia. i.e. the character of the volatility clustering, truncated Levy distribution and scaling, and multifractality of financial time series data present. These are the ground for us to walk further.

## Acknowledgement:

This research is supported financially by Lembaga Pengembangan Fisika Indonesia. Secondly, writer would like to thank for the discussion and critics on the raw draft of this paper in the circle of *Board of Science* Bandung Fe Institute, especially to Yun Hariadi for the discussion of multifractality. The writer also would like to thank to Yohanis for his endeavor in providing data.